\documentclass[12pt,a4paper]{article}

  \usepackage{a4wide}
  \usepackage{latexsym}
  \usepackage{epsf}
  \usepackage{amssymb}
  \usepackage{graphicx}
  \usepackage{amsmath, cite}
 \usepackage{bbm}
  \usepackage{amsmath,amssymb,amsthm}
  \usepackage{verbatim}
  \usepackage{hyperref}
  \usepackage{bbold}
\renewcommand{\d}{\textrm{d}}
\newcommand{\e}{\textrm{e}}

\newcommand{\w}{\wedge}

\newcommand{\SL}{\mathop{\rm SL}}
\newcommand{\SO}{\mathop{\rm SO}}
\newcommand{\SU}{\mathop{\rm SU}}
\newcommand{\be}{\begin{equation}}
\newcommand{\ee}{\end{equation}}
\newcommand{\ba}{\begin{eqnarray}}
\newcommand{\ea}{\end{eqnarray}}
\newcommand{\nn}{\nonumber}
\newcommand{\lp}{\left(}
\newcommand{\rp}{\right)}

\newcommand{\sgn}{\textrm{sgn}\,}

\newcommand\varpm{\mathbin{\vcenter{\hbox{%
  \oalign{\hfil$\scriptstyle+$\hfil\cr
          \noalign{\kern-.3ex}
          $\scriptscriptstyle({-})$\cr}%
}}}}

\newcommand\varmp{\mathbin{\vcenter{\hbox{%
   \oalign{\hfil$\scriptstyle-$\hfil\cr
           \noalign{\kern-.3ex}
          $\scriptscriptstyle({+})$\cr}%
}}}}

\newcommand\varbs{\mathbin{\vcenter{\hbox{%
   \oalign{\hfil$\scriptstyle>$\hfil\cr
           \noalign{\kern-.3ex}
          $\scriptscriptstyle({<})$\cr}%
}}}}

\newcommand\varsb{\mathbin{\vcenter{\hbox{%
   \oalign{\hfil$\scriptstyle<$\hfil\cr
           \noalign{\kern-.3ex}
          $\scriptscriptstyle({>})$\cr}%
}}}}

\begin{document}
\numberwithin{equation}{section}
\begin{flushright}
\small UUITP-15/11\\
\small ITP-UH-03/11\\

\date \\
\normalsize
\end{flushright}
\vspace{0.4cm}
\begin{center}

{\LARGE \bf{The  problematic backreaction of    \\
\vspace{0.3cm}  SUSY-breaking branes
}} \\

\vspace{1 cm} {\large  Johan Bl{\aa}b\"ack$^\dagger$, Ulf
H.~Danielsson$^\dagger$, Daniel Junghans$^\ddagger$, \\
\vspace{0.2cm}Thomas Van Riet$^\dagger$, Timm Wrase$^\sharp$ and Marco Zagermann$^\ddagger$}\\

\vspace{.8 cm}  {${}^\dagger$ Institutionen f{\"o}r fysik och
astronomi\\
Uppsala Universitet, Box 803, SE-751 08 Uppsala, Sweden}\\
\vspace{0.2cm} {\upshape\ttfamily johan.blaback, ulf.danielsson,
thomas.vanriet@fysast.uu.se} \\

\vspace{0.4 cm}  {${}^\ddagger$ Institut f{\"u}r Theoretische Physik \&\\
Center for Quantum Engineering and Spacetime Research\\
Leibniz Universit{\"a}t Hannover, Appelstra{\ss}e 2, 30167
Hannover, Germany}\\

\vspace{0.2cm} {\upshape\ttfamily daniel.junghans,
marco.zagermann@itp.uni-hannover.de} \\

\vspace{0.4 cm}  {${}^\sharp$ Department of Physics, Cornell University, Ithaca, NY 14853, USA}\\

\vspace{0.2cm} {\upshape\ttfamily timm.wrase@cornell.edu} \\

\vspace{1cm}

{\bf Abstract}
\end{center}

\begin{quotation}

\noindent In this paper we investigate the localisation of SUSY-breaking
branes which, in the smeared approximation, support specific non-BPS
vacua. We
show, for a wide class of boundary conditions, that there is no flux
vacuum when the branes are described by a
genuine delta-function. Even more, we find that the smeared solution
is the unique solution with a regular brane profile. Our setup
consists of a non-BPS $AdS_7$ solution in massive IIA supergravity
with smeared anti-$D6$-branes and fluxes T-dual to ISD fluxes in IIB
supergravity.

\end{quotation}

\newpage

\tableofcontents

\section{Introduction}

$D$-branes and orientifold planes are necessary ingredients in
semi-realistic compactifications of type II string theories, where
they play an important role for the implementation of chiral matter,
supersymmetry breaking or tadpole cancelation in the presence of
fluxes. Being massive and charged localised objects, they generally
also backreact on the geometry and source a non-trivial profile
for the warpfactor, dilaton and some of the RR-potentials. Apart
from the special cases where this backreaction is absent due to a
local cancelation of charge and tension for suitably stacked
$D$-branes and $O$-planes, one needs to take into
account these backreaction effects, or make sure they can be
neglected.

The GKP setup \cite{Giddings:2001yu} and its T-duality relatives
(see e.g. \cite{Grana:2006kf,Blaback:2010sj}) are the few examples
where such backreaction effects are quite well
understood\footnote{See \cite{Burgess:2010zt,Burgess:2011mt} for
fully backreacted solutions of codimension-2 branes in 6D
supergravity.}, but much less is known about more general cases,
e.g. with intersecting branes or cases for which the sources and
fluxes are not all mutually BPS as in de Sitter vacua.

A common approach to deal with the effects of localised
sources is to take them into account only in an averaged or
integrated sense. At the level of the 10D field equations, this procedure
corresponds to ``smearing'' the $D$-branes and $O$-planes in the
directions perpendicular to their world volumes\footnote{We refer to
\cite{Koerber:2007hd} for a proper treatment of smeared sources in
flux compactifications. }; in other words, one replaces the
delta-function of the energy and charge density along the transverse
brane coordinates by a constant in
the equations of motion,
\begin{equation}\label{smearing}
\delta \rightarrow \textrm{const}\,.
\end{equation}

On a torus, the smearing could be viewed as a truncation of the
Fourier expansion of the delta-function to the constant term. For
compactifications on more general group or coset manifolds
(including the much studied examples of spheres and (twisted) tori),
this smearing of the sources allows a consistent truncation of the
effective lower-dimensional theory to the sector of left-invariant
modes \cite{Grana:2006kf,Caviezel:2008ik,Cassani:2009ck}.
It is this consistent truncation property
that is usually (albeit often implicitly) exploited in the
derivation of effective gauged supergravity theories, see e.g.
\cite{Roest:2004pk} and references therein. At the level of the 10D
field equations, the smearing in general leads to substantial
simplifications regarding the profile of the warp factor, the
dilaton and some RR-potentials and allows the construction of many
explicit 10D solutions, see e.g. \cite{Acharya:2006ne,Koerber:2008rx}.

As the smearing of a brane-like source is a drastic modification of
the original setup, one should wonder to what extent this can
actually be a good approximation to  a solution with truly localised
branes. More precisely, there are two questions one may ask in this
context:
\begin{enumerate}
\item Does a smeared solution always imply a localised solution, or could
there be smeared solutions that have no well-defined localised
counterpart?
\item In case a smeared solution does have a localised
version, how physically different are these two solutions, e.g.
regarding the vacuum expectation values or masses of stabilised
moduli or the value of the effective cosmological constant?
\end{enumerate}

In this paper we want to address the first question,\footnote{The second
question is what ``warped effective field theory'' is concerned
with, see e.g. \cite{Giddings:2005ff, Koerber:2007xk, Shiu:2008ry, Frey:2008xw, Douglas:2009zn,
Underwood:2010pm} and references therein.}
building upon our previous work \cite{Blaback:2010sj}. There we
studied this question for BPS-type solutions that generalize the GKP
solutions \cite{Giddings:2001yu} to various spacetime dimensions and
found that the BPSness played a crucial role in achieving
successful localisation. Intuitively, this may be understood as a
consequence of the cancelation of forces between the fluxes and the
localised sources in such setups. By BPS-type solutions we mean that
the solutions satisfy a Bogomol'nyi bound, but are not necessarily
supersymmetric. As an example, the GKP solutions
\cite{Giddings:2001yu} are BPS but only supersymmetric when the ISD
flux is of the specific complexity type $(2,1)$ (see also
\cite{Lust:2008zd}).

For non-BPS solutions, on the other hand, it becomes unclear whether
localisation works and simple arguments presented in
\cite{Blaback:2010sj} indicate that it may not work in general. In
this paper we further improve on the arguments given in
\cite{Blaback:2010sj} and consider a particular setup where the
effects of the brane localisation can be traced explicitly.
Concretely, we consider  spacetime-filling (anti-)$D6$-branes on
$AdS_7\times S^{3}$,\footnote{The sphere geometry is not necessary for the smeared solution
as any positively curved Einstein space could fulfill the smeared
equations of motion. The sphere is chosen so as to make a fully explicit discussion of the localised case feasible.} for which a solution exists in the
limit of smeared sources \cite{Blaback:2010sj}. This example
is closely related to the (smeared) Minkowski solutions of GKP
\cite{Giddings:2001yu} and their generalisations to different
dimensions \cite{Blaback:2010sj} and configurations with mixed
sources (cf. App. \ref{app_solutions}). We show that this solution
is perturbatively stable in the left-invariant closed string modes,
which is the sector usually considered in lower-dimensional
supergravity analyses.

Next we study the localisation of the branes. We start with
considering the branes to be fully localised as delta-functions. Under the
assumption that the fluxes very close to the sources are mutually BPS with the source, we
find that the equations of motion do not allow for any possibility
to match the behaviour of the fields, close to and away from the
sources. Therefore it is not possible to construct fully localised
solutions with this BPS-like behaviour near the branes. We point out a caveat in the case that one uses boundary conditions
for which the fluxes near the sources are not mutually BPS with the sources.
However, it is unclear how such boundary conditions could be consistent
with a static solution, since there should be a non-zero force between the fluxes and the sources.
We will present a detailed discussion of boundary conditions at the
sources and a generalisation of our argument to different setups in \cite{toappear}.

In this paper we then present, instead, an independent calculation that supports the above conclusions.
In order to do so, we consider regularised brane profiles that may approximate the previously considered
delta-type profiles to arbitrary precision. If we assume a smooth profile, we find a one parameter family of solutions which correspond
to exactly the one parameter family of coordinate transformations of
the fully smeared solution. This means that there exists no
solutions with regular brane profiles other than the fully smeared
one. This is in contrast to the BPS solutions given by the GKP
solutions and their T-duals, since they solve the equations of
motion for any brane profile. We then use this to rule out also
regularised brane profiles with the shape of step functions.

In the discussion we emphasise the
possible implications of our results on known solutions that are
utilising the smeared approximation of $D$-branes or $O$-planes.

\section{A simple non-BPS setup on $AdS_7 \times S^3$}

In this section we present a simple non-BPS setup on $AdS_7 \times
S^3$ with spacetime-filling (anti-)$D6$-branes as sources. This setup
yields a smeared solution and allows to explicitly tackle the
questions raised in the introduction.

We should point out that this example is a special case of a more
general class of setups for which smeared solutions exist (see App.
\ref{app_solutions}). These setups have $d$-dimensional AdS,
Minkowski or dS spacetimes and a mixture of spacetime filling sources
that allows net charge/tension ratios that can be non-extremal,
 \(T = \eta Q\) with arbitrary \(\eta\). Those solutions with
\(|\eta|=1\) were already discussed in \cite{Blaback:2010sj}. The
specific AdS$_4$ solution with $|\eta|=1$ and $S^3 \times S^3$ as internal
space was first established in \cite{Silverstein:2004id}.

In the following, we shall restrict our discussion to the AdS$_7$
example with \(|Q| = T>0\), since it is the simplest case of this
class of models and we consider it sufficient for supporting the
main point of this paper, namely that localisation of smeared
non-BPS solutions may fail.

\subsection{Type IIA supergravity and smeared $D6/\overline{
D6}$-branes}
\label{sec_smeared}

To establish our notation and conventions we present the equations
of motion for type IIA supergravity with spacetime-filling $D6/\overline{
D6}$-branes in Einstein frame. We use the same conventions as in
\cite{Blaback:2010sj}.\footnote{They are related to the conventions
of \cite{Koerber:2010bx} by going to Einstein frame and changing the
sign of $H$.} Throughout the paper $a,b$ are 10D indices, $\mu, \nu$
are external and  $i,j$ are internal indices. The bosonic
sector of type IIA supergravity contains the metric $g_{ab}$, the
dilaton $\phi$, the $H$ field strength, as well as the RR field
strengths $F_0, F_2, F_4$. Since we compactify to AdS$_7$, however,
we do not consider any external fields that break translational
invariance and hence do not turn on $F_4$.

The trace reversed Einstein equation is then
\begin{align}
\label{eoms_einstein} R_{ab} = & \tfrac{1}{2} \partial_a \phi
\partial_b \phi + \tfrac{1}{2} \e^{-\phi} |H|^2_{ab}-\tfrac{1}{8}
\e^{-\phi} g_{ab} |H|^2 + \tfrac{1}{16} \e^{\tfrac{5}{2} \phi} g_{ab} F_0^2 \\ \notag &{}
+ \tfrac{1}{2} \e^{\tfrac{3}{2} \phi} |F_2|^2_{ab} -\tfrac{1}{16}
\e^{\tfrac{3}{2} \phi} g_{ab} |F_2|^2 +\tfrac{1}{2}(T^{loc}_{ab} - \tfrac{1}{8}g_{ab}T^{loc}),
\end{align}
where $|A|^2_{ab} \equiv \tfrac{1}{(p-1)!}\,A_{a a_2\ldots
a_p}A_{b}^{\,\,a_2\ldots a_p}$, $|A|^2 \equiv
\tfrac{1}{p!}\,A_{a_1\ldots a_p}A^{a_1\ldots a_p}$.

The non-vanishing part of the local stress tensor is given by
\begin{equation}
T_{\mu\nu}^{loc} = - \e^{\tfrac{3}{4}\phi} \mu_6 g_{\mu\nu}
\delta(D6),
\end{equation}
where $\mu_6$ is a positive number, and $\delta(D6)$ is the delta distribution
with support on the $D6$-brane world volume(s), i.e. it may
implicitly also include sums of parallel $D6$-branes. The dilaton equation of motion
is given by
\begin{equation}
\label{eoms_dilaton} \nabla^2 \phi= - \tfrac{1}{2} \e^{-\phi} |H|^2
+ \tfrac{5}{4} \e^{\tfrac{5}{2} \phi} F_0^2 +\tfrac{3}{4}
\e^{\tfrac{3}{2} \phi} |F_2|^2 + \tfrac{3}{4} \e^{\tfrac{3}{4}\phi}
\mu_6 \delta(D6).
\end{equation}
The Bianchi identities for the field strengths are
\begin{equation}
\label{eoms_bianchi} \d H = 0, \qquad \d F_0  = 0, \qquad \d F_2 = H
F_0 \pm \mu_{6} \delta_{3},
\end{equation}
where the upper sign of the source term is for $D6$-branes and the
lower sign for anti-$D6$-branes.  $\delta_{3}$
is shorthand for the normalized volume 3-form transverse to the
$D6$-brane multiplied by $\delta(D6)$, $\delta_3=\delta(D6)\star_3
1$.
\begin{equation}
\int \delta(D6) \star_3 1 = 1.
\end{equation}
The equations of motion for $H$ and $F_2$ read
\begin{equation}
\label{eoms_fluxes} \d(\e^{-\phi}\star H) =
-\e^{\frac{3}{2}\phi}\star F_2 \w F_0, \qquad \d \lp
\e^{\tfrac{3}{2} \phi} \star F_2 \rp = 0.
\end{equation}
In the smeared limit we have
\begin{equation}
\delta(D6) \to \frac{1}{V},
\end{equation}
where $V$ is the volume of the space transverse to the branes.

\subsection{The smeared solution}

In the smeared limit, we can set \(F_2 = 0\) and assume that all other
fields are constant. We furthermore assume that the metric
\begin{equation}
\d s^2_{10}= \mathcal{R}_{AdS}^2 \d s_7^2 + \mathcal{R}_{S^3}^2 \d s_3^2,
\end{equation}
is a direct product $AdS_7 \times S^3$ with $\mathcal{R}_{AdS}$ the AdS radius
and $\mathcal{R}_{S^3}$ the radius of the three-sphere and write
\begin{equation}
H = \lambda F_0 \e^{\tfrac{7}{4}\phi}\star_3 1.
\end{equation}
Going through the equations of motions \eqref{eoms_einstein},
\eqref{eoms_dilaton}, \eqref{eoms_bianchi}, \eqref{eoms_fluxes}, one
then finds that all are solved for
\begin{equation}
\mathcal{R}_{S^3}^2=\tfrac{\mathcal{R}_{AdS}^2}{12} = \tfrac{7^{7/8}}{\sqrt{2} \, 5^{5/4}} \lp F_0^2 h^5\rp^{1/4}, \quad \e^{\phi} =
\sqrt{\tfrac{20}{7\sqrt{7}}} \tfrac{1}{\sqrt{F_0^2 h}} , \quad \lambda = \varmp
\frac{5}{2}, \quad \mu_6 = 2 |F_0| h \pi^2,
\end{equation}
where $h = \sqrt{|H|^2 \mathcal{R}_{S^3}^6}$ is independent of $\mathcal{R}_{S^3}$. The (lower) upper sign in $\lambda$ is for
(anti-)$D6$-branes. For large flux parameters, we thus have large volumes and small string coupling.

\section{Perturbative stability}

In this section we want to verify the perturbative stability of the
smeared solution of the previous section. For practical reasons, we
do this only for the subsector of the left-invariant modes. This is
precisely the sector that underlies the usual gauged supergravity
description. In case there happen to be tachyonic modes among the higher
Kaluza-Klein modes or in the sector that is not left-invariant, they would hence also be missed in the gauged
supergravity approach, which is the standard tool to find new vacua.

We first consider the closed string sector and perform the
computation in all details. For the open string sector we are
necessarily sketchy and point out where possible instabilities could
reside, if any.

\subsection{The left-invariant closed string moduli}

When we regard the three-sphere as the group manifold $\SU(2)$ there
is a standard way to take into account a subset of the degrees of
freedom, namely the left-invariant modes \cite{Scherk:1979zr}. The
left-invariant modes are the lower-dimensional fields that one
obtains by expanding the supergravity fields in the left-invariant
Maurer--Cartan basis of $\SU(2)$. These forms form a global coframe
on the group manifold and obey the characteristic equation
\begin{equation}
\d e^i= -\tfrac{1}{2}f^i_{jk} e^j\wedge e^k\,.
\end{equation}
We take a basis for the structure constants for which
\cite{AlonsoAlberca:2003jq}
\begin{equation}
f^i_{jk} = q\,\epsilon_{jkl}\,\delta^{li}\,,
\end{equation}
where $q$ is a number that gives the curvature scale.

There are six metric scalars, three scalars from $C_1$, three
from $B_2$, one from $C_3$ and one being the dilaton $\phi$ in ten
dimensions. This adds up to 14 scalars that form the coset manifold
$\SL(5)/\SO(5)$. Of course when we add the ``open string degrees'' of
freedom, coming from the $D6/\overline{D6}$-branes, we have more scalars, and the
coset enlarges.

Let us write the moduli explicitly in a convenient basis, by which
we mean a field basis for which the field metric, at the solution, is canonically normalised. This then allows us to
compute
the scalar masses by simply diagonalising $\partial_i\partial_j V$.

The 10D metric, in Einstein frame, is written as follows
\begin{equation}
\d s^2_{10} = \e^{2\alpha v}\d s_7^2 + \e^{2\beta v}\d s_3^2,
\end{equation}
where $v$ is the volume modulus and the numbers $\alpha$ and $\beta$
are chosen such that the reduced theory is in Einstein frame and
that the volume modulus is canonically normalised
\begin{equation}
\alpha^2=\frac{3}{80}\,,\qquad \beta=-\frac{5}{3}\alpha\,.
\end{equation}
The internal metric is then written as
\begin{equation}
\d s_3^2 = M_{ij}\,e^i\otimes e^j,
\end{equation}
where $M$ is the symmetric, positive definite, metric-moduli matrix.
It is explicitly given by $M=LL^T$, with $L$ the coset
representative of $\SL(3)/\SO(3)$ in a solvable basis
\begin{equation}
L=\begin{pmatrix}
\e^{\tfrac{1}{2}\sigma_1+\tfrac{1}{2\sqrt{3}}\sigma_2} & \e^{-\tfrac{1}{2}\sigma_1+\tfrac{1}{2\sqrt{3}}\sigma_2}\chi_1 & \e^{-\tfrac{1}{\sqrt3}\sigma_2}(\chi_1\chi_2+\chi_3)\\
 0 & \e^{-\tfrac{1}{2}\sigma_1+\tfrac{1}{2\sqrt{3}}\sigma_2} & \e^{-\tfrac{1}{\sqrt3}\sigma_2}\chi_2\\
 0 & 0 & \e^{-\tfrac{1}{\sqrt3}\sigma_2}.
\end{pmatrix}
\end{equation}
Note that det$(M)=1$. The five scalars, $\sigma_1, \sigma_2, \chi_1, \chi_2$ and $\chi_3$
are canonically normalised and together with the volume modulus $v$
we have six metric scalars as announced previously. The on-shell
values $\sigma_1=\sigma_2=\chi_i=0$ give the ``round sphere'', which
corresponds to our solution.

The $C_3$ field gives one canonically normalised axionic scalar $a$
via $C_3 =  a \epsilon_3$. The $B$-field and $C_1$ field both give
rise to three canonically normalised scalars $b^i$ and $c_i$ as
follows
\begin{align}
&B=\tfrac{1}{2}b^i\,\epsilon_{ijk}\,e^j\wedge e^k\,,\\
&C_1 = c_i e^i\,.
\end{align}
The scalar potential gets contributions from the internal curvature,
$V_R$, the $H$-flux, $V_H$, the RR p-form fluxes, $V_p$, and the
source tension, $V_{D6}$.  We explicitly find
\begin{align}
&V_R = \e^{(7\alpha+\beta)v}\, q^2\,\Bigl(-
\tfrac{1}{2}(\text{Tr}[M])^2+\text{Tr}[M^2] \Bigr)\,,\\
&V_0 = \tfrac{1}{2}\,\e^{\tfrac{5}{2}\phi +
(7\alpha+3\beta)v}\,F_0^2\,,\\
& V_{D6} = \e^{\tfrac{3}{4}\phi + 7\alpha v}\,T_{D6}\,,\\
& V_H = \tfrac{1}{2}h^2 \e^{-\phi + (7\alpha-3\beta)v}\,,\\
&V_2 = \tfrac{1}{2}\e^{\tfrac{3}{2}\phi +
(7\alpha-\beta)v} \Bigl(-q c_i\delta^{mi} + F_0 b^m\Bigr) \Bigl(-q
c_j\delta^{nj} + F_0 b^n\Bigr)M_{mn}\,,
\end{align}
where we have taken the $H$-field to be given by
\begin{equation}
H = h\epsilon_3 + \d B,
\end{equation}
with $B$ as above. In order for the $\d B$ term to
generate a mass term for the $b^i$-scalars we need $\d B \sim
\epsilon_3$, which cannot be done given the cohomology of $\SU(2)$.
$F_4$ is not contributing any energy and hence, $a$ is a massless
axion.

Our AdS solution stabilises the combinations $F_0 b^m-q
c_i\delta^{mi}$ while the orthogonal combinations remain flat. This
allows us to set $b^i=c_i=0$. Also, since the three-sphere is round,
we have $M=\mathbbm{1}$, or, equivalently,
$\sigma_1=\sigma_2=\chi_1=\chi_2=\chi_3=0$ at the vacuum. We also choose the
dilaton, $\phi$, the axion $a$ and the volume scalar $v$ to be
zero. Then all scalars are zero at the solution and we are at the
origin of the scalar field space, where the metric is canonically
normalised. These field values imply that we have to take the fluxes and
tension appropriately
\begin{equation}
F_0^2=\tfrac{4}{25}h^2\,,\qquad T=\tfrac{2}{5}h^2\,,\qquad
q^2 = \tfrac{28}{25} h^2\,.
\end{equation}
One can easily check that the origin in field space (i.e. all
scalars zero) is an extremum of the potential.

Mathematica finds the following eigenvalues of $\partial_i\partial_j
V|_{\phi^i=0}$ at the origin of field space
\begin{equation}
\lambda = h^2 \Bigl\{\tfrac{56}{25}, \tfrac{32}{25},\tfrac{2}{25} (18 \pm \sqrt{79}), 0 \Bigr\}\,,
\end{equation}
where the multiplicity of $\lambda=\tfrac{56}{25} h^2$ is $5$, the multiplicity of
$\lambda=\tfrac{32}{25} h^2$ is $3$ and the multiplicity of $\lambda=0$ is $4$.
The other two eigenvalues have multiplicity  equal to $1$. Since there
are no negative mass modes the solution is stable with respect to
these 14 left-invariant degrees of freedom.

\subsection{Open string moduli}

We have not taken into account the open string moduli since this is
technically more challenging. The appropriate technique for this
would be to construct the corresponding half-maximal gauged
supergravity in $D=7$ since the gauged supergravity also contains
the open string fields. We leave this for further research, but
already comment on what one could expect on general grounds.

First of all one naively expects the moduli associated with the
brane position to be massless as the round sphere is homogenous and
hence without preferred points. The same is true for our fluxes that
fill the sphere. However, there is the possibility of brane-flux
annihilation, which can be a perturbative effect in some cases
\cite{Kachru:2002gs}. Note that the fluxes behave as
$D6$-branes when the sources are pure anti-$D6$-branes and vice
versa. Since the amount of fluxes equals the amount of brane charge
the flux/brane annihilation process should be perturbative, at least
this was the case for geometries dual to cascading gauge theories
in \cite{Kachru:2002gs}. It seems sensible that this applies to our
situation as well. Brane/flux annihilation should occur via the
Myers effect \cite{Myers:1999ps}, which implies the nucleation of a
$D8$-brane, wrapping a 2-sphere inside the 3-sphere. However, this
presumed instability would be absent for the solution with just a
single $D6$-brane. It would be interesting to study the decay
further and give a quantitative description. However, in this paper
we want to understand the relation between smeared and localised
sources, which we take to be a closed string problem.

\section{Localisation}

In \cite{Blaback:2010sj} an intuitive argument was presented explaining why it is difficult to
find static configurations composed of mutually non-BPS components. A
simple example is an anti-brane sitting in a background of ISD flux. What will
happen in general is that the flux is attracted to and sucked up by the brane,
and possibly annihilated. The equations that capture this behaviour in our setup are
primarily the $F_2$ Bianchi identity and the equation of motion for the $H$ field.
The Bianchi identity determines $F_{2}$ in terms of the sources, including
$F_0 H$, and the equation of motion for $H$ determines how $H$ responds to $F_{2}$.
A combination of both equations provides a very powerful tool to
restrict possible configurations.

In the following we will treat two cases, first we consider the
case of delta-function sources and find strong constraints
that rule out a large class of configurations. Then we focus on extended
and fully regularised source profiles. We analyse
the full system using a Taylor expansion, where we find some
surprisingly strong analytical constraints on what kind of
configurations are possible. Our results show that the fully
smeared solution is the only regular solution to the equations of
motion.

\subsection{The ansatz}

As a starting point we assume that all sources are located on the
north or/and south pole of the three-sphere such that the setup
preserves an $\SO(3)$-symmetry of the full $\SO(4)$ symmetry of the
smeared solution. Although the backreaction might bring us away from
the three-sphere, we see no physical reason for why it should break
the symmetry exhibited by the setup. We can then assume that for any
setup preserving these isometries one can find a metric which at
most has the physical singularities
(i.e. the sources) at the poles. The most general form of such a metric is
\begin{equation}
\d s^2_{10}=\e^{2A(\theta)}\d s_7^2 +\e^{2B(\theta)}\Big( \d\theta^2
+\e^{2C(\theta)} \sin^2(\theta)\d\Omega_2^2 \Big),
\end{equation}
where the second conformal factor \(C\) can be absorbed into \(B\)
by a suitable coordinate transformation \(\theta \to \tilde \theta
(\theta)\) that keeps the sources at the poles. One can show that,
if the original metric does not have any singularities away from the
poles, then the same is true for the new metric for which $C$ is transformed
away. Hence, without loss of generality, we can write
\begin{equation}
\d s^2_{10}=\e^{2A(\theta)}\d s_7^2 +\e^{2B(\theta)}\Big(\d\theta^2 +
\sin^2(\theta)\d\Omega_2^2\Big),
\end{equation}
and assume \(A,B\) to be regular away from the poles.

The flux ansatz is
\begin{align}
& H=\lambda F_0 \e^{\tfrac{7}{4}\phi}\star_3 1\,, \label{eq:Hflux}\\
& F_2 = \e^{-\tfrac{3}{2}\phi-7A}\star_3\d\alpha\,,
\end{align}
where $\phi, \lambda$  and $\alpha$ are now functions depending on
$\theta$ and we take $F_0$ to be constant. This is the most general
ansatz compatible with the form equations of motion
\eqref{eoms_bianchi} and \eqref{eoms_fluxes} and our symmetries. In
our previous work \cite{Blaback:2010sj}, we have considered a similar
ansatz for the AdS$_4\times S^3\times S^3$ solution, but did not
take into account a non-trivial profile for the variable $\lambda$. We then found an obstacle
to localisation. Allowing the variable $\lambda$ is the most general
ansatz and we should be able to fully settle the discussion on
whether localisation is possible or not.

Hence, the problem is reduced to finding a set of five unknown functions $A, B,
\phi, \lambda, \alpha$ depending on $\theta$ and obeying coupled
second-order differential equations, which we now derive. A prime
$'$ denotes the derivative with respect to $\theta$, e.g., $A'=\d A/\d
\theta$.

The $F_2$-Bianchi identity implies
\begin{equation}
\frac{\Bigl(\e^{-\tfrac{3}{2}\phi-7A+B}\sin^2\theta
\alpha'\Bigr)'}{\e^{3B}\sin^2\theta}=\e^{\tfrac{7}{4}\phi}\lambda
F_0^2 +Q\delta(D6)\,, \label{eom1}
\end{equation}
where $Q\delta(D6)$ symbolizes all
source contributions at the north and south pole and is therefore
generically a sum of delta-functions. The $H$
equation of motion allows us to eliminate $\alpha$ in terms of
$\lambda$
\begin{equation}\label{alphaversuslambda}
\alpha=\e^{\tfrac{3}{4}\phi +7A}\lambda +\alpha_0\,.
\end{equation}
The dilaton equation gives
\begin{equation}
\frac{\Bigl(\e^{7A+B}\sin^2\theta
\phi'\Bigr)'}{\e^{7A+3B}\sin^2\theta}=\e^{\tfrac{5}{2}\phi} F_0^2
\bigl(\tfrac{5}{4}-\tfrac{\lambda^2}{2} \bigr) + \tfrac{3}{4}\e^{-
14A - 2B - \tfrac{3}{2}\phi}(\alpha')^2 +
\tfrac{3}{4}\e^{\tfrac{3}{4}\phi} T\delta(D6)\,, \label{eom2}
\end{equation}
where $T\delta(D6)$ symbolizes the sum over all localised
tension contributions. The trace of the external Einstein equation gives
\begin{align}
\tfrac{16}{7} R_7 =& -96\e^{-2A}  -16 \e^{-2B} \lp 7(A')^2 + A'B'
+ \frac{(\sin^2\theta A')'}{\sin^2\theta} \rp\nonumber\\
=& \, \e^{\tfrac{5}{2}\phi} F_0^2 (1-2\lambda^2) -
\e^{-14A-2B-\tfrac{3}{2}\phi}(\alpha')^2  -
\e^{\tfrac{3}{4}\phi}T\delta(D6)\,. \label{eom3}
\end{align}
The internal Einstein equation is more involved. We first display
the Ricci tensor
\begin{align}
& R_{\theta\theta} = 2  - \frac{(\sin^2\theta B')'}{\sin^2\theta} - 7(A')^2 - B'' -7A'' + 7A'B'\,,\\
& R_{\varphi \varphi} = \e^{-2B} g_{\varphi \varphi} \lp 2  - (B')^2 -\frac{(\sin^2\theta
B')'}{\sin^2\theta} -\cot\theta(B+7A)' -7 A'B' \rp \,,
\end{align}
where $\varphi$ indicates the two remaining angles. The
Einstein equation in the ($\theta \theta$)-direction leads to
\begin{align}
&  2  - \frac{(\sin^2\theta B')'}{\sin^2\theta} - 7(A')^2 - B''
-7A'' + 7A'B' = \nonumber\\
& \tfrac{1}{2}(\phi')^2 + \tfrac{1}{16}\e^{\tfrac{5}{2}\phi + 2
B}\,F_0^2\Bigl(1+6\lambda^2\Bigr)-\tfrac{1}{16}\e^{-14A-\tfrac{3}{2}\phi}(\alpha')^2
+\tfrac{7}{16}\e^{\tfrac{3}{4}\phi +2B}T\delta(D6)\,. \label{eom4}
\end{align}
whereas the Einstein equations in the ($\varphi \varphi$)-directions lead to
\begin{align}
& 2  - (B')^2 -\frac{(\sin^2\theta B')'}{\sin^2\theta}
-\cot\theta(B+7A)'
-7 A'B' = \nonumber\\
& \tfrac{1}{16}\e^{\tfrac{5}{2}\phi + 2
B}\,F_0^2\Bigl(1+6\lambda^2\Bigr) +
\tfrac{7}{16}\e^{-14A-\tfrac{3}{2}\phi}(\alpha')^2 +
\tfrac{7}{16}\e^{\tfrac{3}{4}\phi +2B}T\delta(D6)\,. \label{eom5}
\end{align}

\subsection{Delta-function sources}

In the case of delta-function sources our tool is the combination of
the $H$ equation of motion and the Bianchi identity for $F_2$. At
any point away from the sources we can combine
(\ref{eom1}) and (\ref{alphaversuslambda}) by substituting $\lambda$
to find
\begin{equation}
\frac{\Bigl(\e^{-\tfrac{3}{2}\phi-7A+B}\sin^2\theta
\Bigr)'}{\e^{3B}\sin^2\theta}\alpha' + \e^{-\tfrac{3}{2}\phi-7A -2B}
\alpha''=\alpha \e^{\phi-7A} F_0^2\,,
\end{equation}
where we have used that through a shift in $\alpha$ we can
always set $\alpha_0=0$.

This equation tells us one important fact: For non-singular $\e^\phi,\e^A,\e^B$, the function
$\alpha$ must obey at any extremum ($\alpha'=0$)
\begin{equation}\label{constraint}
\boxed{\sgn \alpha'' = \sgn \alpha\,.}
\end{equation}
Note that this is also true at either of the poles ($\theta =0,\pi$), if there are no localised sources at that pole.
To be able to make use of this equation, we need to know the behaviour of $\alpha$
close to the (anti-)brane. This brings us to the difficult
discussion of what the boundary conditions should be, resemblant of
the discussion of the boundary conditions for anti-$D3$-branes at
the tip of the conifold \cite{Bena:2009xk, McGuirk:2009xx}.

\subsubsection*{BPS-like boundary conditions}
Global tadpole cancellation requires the integrated flux to be of opposite BPS-type as the brane itself,
which immediately raises concerns about the possible stability of the localised system.
On the other hand, it is only the integrated flux that is fixed by tadpole cancellation, so one may hope that perhaps a judicious distribution of the flux relative to the branes and a suitable choice of boundary conditions could result in a static solution.

The simplest possibility in this respect would be
 to take the boundary conditions as
defined by the BPS $D6$-brane solution in massive IIA supergravity \cite{Janssen:1999sa}, which also
involves H-flux and the Roman's mass $F_0$, as well as $F_2$-flux magnetically sourced by the brane.
With these boundary conditions, the flux near the brane would be BPS with respect to the brane itself, and one would at least
not expect instabilities to occur from the region close to the brane. As one moves away from the brane,
the flux gradually has to change its BPS-type, as required by tadpole cancellation. Whether such a configuration has a chance to be also globally stable is not at all obvious. The interesting point now is that we actually do not need to know anything about the global stability of this flux setup, since we can rule it out completely using a simple `topological' argument based on (\ref{constraint}).

\begin{figure}[h!]
\centering
\includegraphics[width=0.4\textwidth]{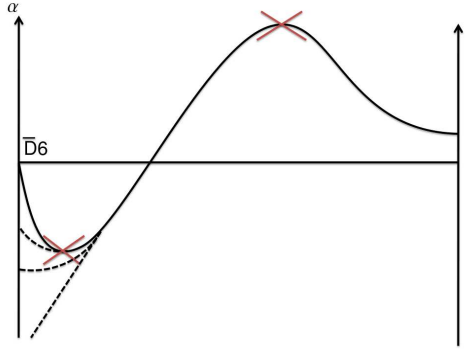}
\caption{Configuration of a single $\overline{D6}$-brane and flux.  The solid line corresponds to the usual boundary condition
of a BPS $\overline{D6}$-brane, whereas the dashed lines represent more general boundary conditions consistent with net `IASD' flux near the $\overline{D6}$-brane. Global tadpole cancellation then enforces the forbidden extrema marked by the crosses. \label{fig:1aD6}}
\end{figure}

\begin{figure}[h!]
\centering
\includegraphics[width=0.4\textwidth]{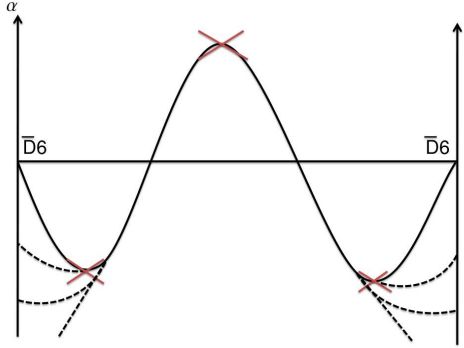}
\caption{A configuration with a $\overline{D6}$-brane at each pole with assumed `IASD' flux near the branes,
 again leading to forbidden extrema. \label{fig:2aD6}}
\end{figure}

To see this, let us literally assume the standard BPS-boundary conditions at the (anti-) $D6$-brane\footnote{The main differences of the BPS solutions \cite{Janssen:1999sa} to our setup are the different world volume geometries (AdS vs. Minkowski), the compactness and curvature of the transverse dimensions and the different sign of the integrated flux $HF_0$. Zooming into the brane region with an assumed  locally BPS flux, however, the large scale curvature and global structure should not
matter.} so that we have a near horizon region at the brane, where $\e^A$ and $\e^\phi$ approach zero, so that, in particular, $\e^{\tfrac34\phi+7A} \rightarrow 0$.
An assumed `I(A)SD' flux\footnote{In lack of a better word, we will call the flux for $\lambda =\varpm 1$ that is mutually BPS with $O6$-planes and $D6$-branes ($\overline{O6}$-planes and $\overline{D6}$-branes) `I(A)SD' and flux with $F_0 H = \alpha F_0^2 \e^{\phi-7A} \star_3 1$ with $\alpha \varbs 0$ net `I(A)SD'.}
near the (anti-)brane would mean $\lambda = \varpm 1$, and hence, using \eqref{alphaversuslambda},
$\alpha \varbs 0$ for an (anti-)brane starting at $\alpha=0$. In Figures \ref{fig:1aD6} and \ref{fig:2aD6}, these standard BPS boundary conditions lead to the $\alpha$ profile indicated by the solid lines.
If we do not insist on the flat space boundary conditions but still require a net `I(A)SD' flux near the (anti-)brane, we would still have that $\alpha \varbs 0$ near the brane, corresponding to the dashed lines in Figure \ref{fig:1aD6} and \ref{fig:2aD6}.

The important point now is that, to ensure  tadpole cancellation, the total \emph{integrated} flux for (anti-)branes has to have exactly the opposite sign,
$\int F_0 H = F_0^2 \int \alpha \, \e^{\phi-7A} \star_3 1 \varsb 0$. Hence $\alpha$ must at some point change its sign. This
generically requires an extremum with $\sgn \alpha'' = -\sgn \alpha$, in contradiction with (\ref{constraint}),
so that with the BPS-like boundary conditions considered here, we can easily rule out certain
combinations of fluxes and sources that are not mutually BPS.

Let us illustrate this with two examples. First consider the case of a single anti-D6-brane located at $\theta=0$
and no source at $\theta=\pi$. This case is depicted in figure
\ref{fig:1aD6}.

Starting from the anti-brane, we have, close to
the brane, $\alpha < 0$, since the flux is locally net `IASD'. To cancel the tadpole,
the integrated flux must be positive, and hence $\alpha$ must become
positive somewhere between the poles. Since we also have to have an extremum at the pole at $\theta =\pi$ due to rotational symmetry, $\alpha$ is forced to at least one extremum that violates
(\ref{constraint}) (marked by crosses in the figure). Note that, due to \eqref{constraint}, the local maximum in Figure \ref{fig:1aD6} would also be excluded if it happens to be at the pole without the brane.
We can also consider this case starting from the point
$\theta=\pi$, where, as mentioned above, we have an extremum due to the rotational symmetry and the absence of a source.
If, as is the case depicted in the figure, we start out at some positive $\alpha$ at that pole, $\alpha$ must
grow. Then, to match with the boundary condition of the anti-brane at $\theta=0$,
$\alpha$ has to have a maximum with $\alpha>0$, which is again forbidden
by (\ref{constraint}). If we start out at some negative
$\alpha$ at $\theta=\pi$, $\alpha$ must decrease, and we could potentially match
with the anti-$D6$-brane at $\theta=0$. However, the net flux charge $\int H F_0$ would then be negative
so that now the tadpole constraint would be violated. The case $\alpha(\pi)=0$
can be excluded in a similar manner.

In the second example, we place one anti-brane at $\theta=0$ and another at
$\theta=\pi$. This situation is depicted in figure \ref{fig:2aD6}.
Starting at $\theta=0$ the situation is just as before, we need
$\alpha$ to flip sign to be able to cancel the tadpole, which leads to an extremum violating
(\ref{constraint}). In this case, coming from $\theta=\pi$
is just the mirrored situation. That is, it is not possible to match
the `IASD' boundary conditions of two anti-branes.

\begin{figure}[h!]
\centering
\includegraphics[width=0.4\textwidth]{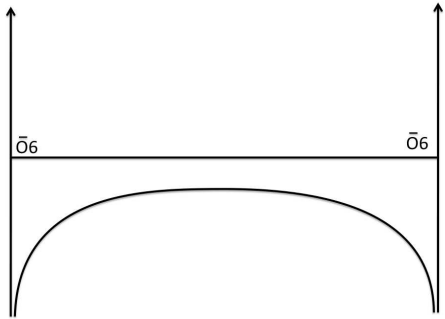}
\caption{BPS configuration with orientifold-planes.\label{fig:bps1}}
\end{figure}

One might wonder how BPS setups constructed of anti-orientifold planes
and `IASD' fluxes fit in our discussion. Although the geometry
is different in those cases, our qualitative argument can be applied similarly, since
it does not rely on the details of the internal space. We show a setup with
anti-$O6$-planes at each pole in figure \ref{fig:bps1}.

The boundary conditions for the anti-$O6$-planes are
different from the standard $D6$-brane boundary conditions: instead of having $\e^{\tfrac34 \phi+7A}$ zero, this factor diverges
at the plane, hence $\alpha$ diverges. Furthermore, the
anti-orientifold charge and the `IASD' flux are already mutually BPS,
and $\alpha$ never needs to switch sign to cancel the tadpole. This
is the reason why the BPS situations survive localisation whereas the non-BPS
situations collapse.

\subsubsection*{Alternative boundary conditions}

\begin{figure}[h!]
\centering
\includegraphics[width=0.4\textwidth]{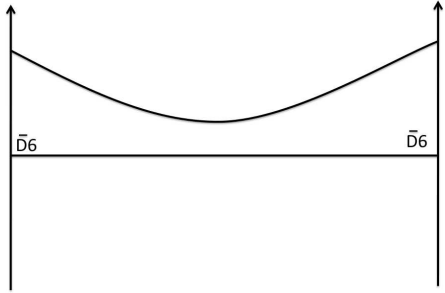}
\caption{Alternative boundary conditions\label{Figure5}.}
\end{figure}

Obviously one can draw a curve for $\alpha$, consistent
with the global tadpole cancelation, that obeys $\eqref{constraint}$
in its extrema. One example with an equal number of
anti-$D6$-branes at the north- and south-pole is given in figure
\ref{Figure5}.

Note however that when the warp factor and the dilaton go to zero near the source, our argument rules out all net `ISD' boundary conditions with finite
$\lambda$. The only net `ISD' cases we cannot rule out with this argument would then be such that $\lambda$ blows up at least as fast as
the warp factor and dilaton go to zero so that $\alpha$ does not vanish (cf. \eqref{alphaversuslambda}). Furthermore, the boundary conditions would have to be such that $\alpha' <0$ at the anti-$D6$-brane. It would be quite surprising if  
such solutions could be static, since the
fluxes near the branes are not mutually BPS with the sources, so that
one would expect net forces between the branes and the fluxes.
It should be
possible to check explicitly whether these anti-BPS-like boundary conditions
could make any sense at all in our model, because the boundary
conditions have to be consistent with the tension and charge
multiplying the delta-function of the sources, which excludes many possible
boundary behaviours. We will discuss this in more detail in \cite{toappear}.

In the rest of this paper, we instead give another argument against the localisability
of the smeared solution presented in section 3, by viewing perfectly localised sources
as a limiting case of regularised brane profiles of finite extent. This argument will be very general and does not
use stability considerations or assumptions about boundary conditions and can hence be viewed as a
complementary confirmation of the findings of the present subsection.

\subsection{Regularised sources}

We now explore the possibility of solutions with a regularised source
profile, where the source is non-zero at points away from the poles.
Let us consider extremal sources \(|Q|=T>0\) that have a regular
profile with the source term written as
\begin{equation}\label{sourceterm}
T\delta(D6) = z(\theta) \e^{-3B}\,,
\end{equation}
with $z(\theta)$ being a smooth function that becomes
a constant in the smeared limit. The reason we took out the
$\e^{-3B}$ factor is that the delta-function, as it appears in the
equations of motion, contains an inverse determinant of the metric.

Our setup has a consistent solution, if we can solve the equations
of motion \eqref{eom1}, \eqref{eom2}, \eqref{eom3}, \eqref{eom4},
\eqref{eom5}. We can check the consistency of these five
differential equations by means of a Taylor expansion
\ba
&\e^{2A(\theta)} = \sum \limits_{n} a_n (\theta-\theta_0)^n, \qquad
&\e^{2B(\theta)} = \sum \limits_{n} b_n (\theta-\theta_0)^n, \nn \\
&\e^{\tfrac14 \phi(\theta)} = \sum \limits_{n} f_n
(\theta-\theta_0)^n, \qquad &\lambda(\theta) = \sum \limits_{n}
\lambda_n (\theta-\theta_0)^n, \ea around an arbitrary point
\(\theta_0 \in ]0,\pi[\), where \(z(\theta_0) \neq 0\). Plugging
this expansion into the equations of motion, their coefficients \(\sim
(\theta-\theta_0)^n\) yield algebraic equations for every order
\(n\) that have to be satisfied at \(\theta=\theta_0\). At every
order \(n\), one can then use the four equations \eqref{eom1},
\eqref{eom2}, \eqref{eom3}, \eqref{eom4} to determine \(a_{n+2},
b_{n+2}, f_{n+2}, \lambda_{n+2}\) in terms of the lowest order
coefficients \(q=\left\{a_0, b_0, f_0, \lambda_0, a_1, b_1, f_1,
\lambda_1\right\}\). The remaining fifth equation \eqref{eom5} then
gives a constraint between the elements of \(q\) only. At zeroth
order (\(n=0\)), for example, \(a_2, b_2, f_2, \lambda_2\) are
determined in terms of \(q\), whereas the fifth equation yields an
additional constraint for \(q\), and so on. Going to higher orders
\(n\), potentially produces an infinite amount of constraints between
the elements of \(q\) and thus should generically lead to strong
constraints.\footnote{This operation of finding constraints is very
time consuming and sometimes, if not done in the correct order, too
complex to solve on a desktop computer. Therefore the reader
interested in verifying these computations should contact the
authors to get more precise instructions.}

From the equations of the first six orders, we get the following
conditions (upper sign for $D6$-branes, lower sign for anti-$D6$-branes
as sources)
\begin{equation}
a_1 = f_1 = \lambda_1 = 0, \qquad \lambda_0 = \mp \tfrac{5}{2},
\qquad F_0^2 = \tfrac{48}{7} \tfrac{1}{a_0 f_0^{10}}.
\end{equation}
Since \(\theta_0\) was arbitrary
these constraints must be satisfied for all $\theta_0 \in ]0,\pi[$, and hence
\begin{equation}
\e^{2A(\theta)} = a = \mathrm{const.}, \quad \e^{\tfrac14 \phi(\theta)} =
f = \mathrm{const.}, \quad \lambda(\theta) =  \mp \tfrac{5}{2},
\quad F_0^2 = \tfrac{48}{7} \tfrac{1}{af^{10}},
\end{equation}
everywhere (except at the poles). We can then plug this ansatz into the equations of
motion \eqref{eom1}, \eqref{eom2}, \eqref{eom3}, \eqref{eom4},
\eqref{eom5} and solve for \(B(\theta)\) and \(z(\theta)\). We find
that all equation are solved for
\begin{equation}
\e^{2B(\theta)} = \tfrac43 \left[{\xi \cos(\theta)-\sqrt{\tfrac{16}{a}
+\xi^2}}\right]^{-2}, \qquad z(\theta) = \tfrac{320}{7 \sqrt{3}\,a f^3}
\left|{\xi \cos(\theta)-\sqrt{\tfrac{16}{a}+\xi^2}}\right|^{-3}.
\label{eq:a}
\end{equation}

This solution has $z(\theta) \e^{-3B}$ constant, which, with (\ref{sourceterm}) and the fact that
all fields except $B$ are constant, suggests that this is just the smeared solution written in a different
coordinate system of the sphere. To find the corresponding coordinate transformation, consider
\begin{equation}
\d s_{S^3}^2 = \mathcal{R}_{S^3}^{2}(\d {\theta'}^2 + \sin^2 \theta' \d \Omega^2) = \e^{2B(\theta)} \left( \d \theta^2 + \sin^2 \theta \d \Omega^2 \right).
\end{equation}
Eliminating $\theta^{\prime}$ in terms of $\theta$ provides an ordinary differential equation for
$\e^{2B(\theta)}$ with solutions which agree exactly with
(\ref{eq:a}). This means that we have proven that the smeared
solution is the only solution to the equations of motion for
which the source profile is a smooth function.

\begin{figure}[h!]
\centering
\includegraphics[width=.4\textwidth]{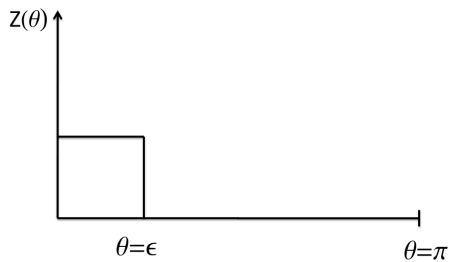}
\caption{A brane source that is a step function.\label{Figure6}}
\end{figure}

Note that our proof for obtaining the profile of $z(\theta)$ was
valid in the support of $z(\theta)$. Hence one could still study
the case for which, instead of an
everywhere smooth function, the source profile is a step function as
in figure \ref{Figure6}.

Then we should solve the vacuum equations outside of the source and
match at the boundary $\theta=\epsilon$. For this to work the
solution outside the source region should have a zero integrated
total flux induced charge
\begin{equation}
\int^{\pi}_{\epsilon}F_0 H =0\,.
\end{equation}
This follows because the regularised brane itself cancels exactly the flux-induced charge that
is within the brane region ($\theta < \epsilon$) so that the region outside
the brane also has to have vanishing charge. Put differently, the source swallows as
much flux within its spatial extent as it has charge. This
implies that $\lambda$ is varying and switches sign outside the
source. However, it is easy to prove that this cannot occur, using
the same reasoning as before based on (\ref{constraint}). The only thing
we need is the behaviour of the fields at the boundary of the
regularised source, where $\theta=\epsilon$. Assuming that the
fields and their first derivatives are continuous at the step, we
must have $\alpha'=0$ (the same is true for the other fields), since
otherwise we cannot consistently connect to the smeared solution
inside the region $\theta<\epsilon$. Take for simplicity the
solution which has positive $\alpha$ at the regularised source and
$\alpha_0=0$ (the latter can be obtained by simply shifting $\alpha$).
Then we have that, at $\theta=\epsilon$, (cf. (\ref{constraint}))
\begin{equation}
\alpha''>0\,.
\end{equation}
So the function $\alpha$ is increasing when it leaves the brane. But
we argued that it has to become negative in order to cancel the
global tadpole. Therefore it has to reach a maximum somewhere.  At
the would-be maximum we have $\alpha'=0$ and hence we would again
find $\alpha''>0$ inconsistent with it being a maximum. Therefore
$\alpha$, and thus $\lambda$, can never switch sign in order to
cancel the global tadpole. The same happens when there are box-like
brane sources at both poles.

In \cite{Blaback:2010sj} we conjectured that a static non-BPS
solution of the above kind could not be localised since the fluxes,
for fixed $\lambda$, behave as smeared branes with the opposite
charge of the real brane that cancels the tadpole. Therefore one
expects the fluxes to be attracted into the localised branes. When
the branes are smeared out this process does not occur since there is
no preferred point of attraction. Rather one would expect brane-flux
annihilation to take place but this is outside of the closed string
sector we consider. One could have expected the no-go of
\cite{Blaback:2010sj} to be evaded by having a varying $\lambda$, in
the same way that anti-branes at the tip of the warped conifold,
locally generate IASD fluxes \cite{DeWolfe:2008zy}, which, far away
from the anti-branes change back into ISD fluxes. However, our above
proof indicates that, when we do this for regularised branes, this
is not possible and $\lambda$ remains constant, and hence the only
solution is the fully smeared one.

\subsection{Summary of results}

In the case of fully localised
sources we have shown, using ``topological'' arguments, that it is
not possible to construct solutions, for a large class of boundary
conditions. These arguments only use the Bianchi identity for $F_2$
and the $H$ equation of motion. The power of the argument is in its
global nature. It could be that solutions exist locally around the
north- and south-pole, but they are bound to collapse somewhere in
the middle. Note that these arguments specifically use the choice of
boundary conditions and the global tadpole cancellation condition. Hence in a
non-compact situation (like the conifold) things might change. But
non-compact models should be regarded as local models that
eventually need an embedding in a global compact model for which the
global tadpole indeed matters. The dependence on the boundary
condition is also essential, and we have shown cases where
there might be an option to evade our no-go for a given set of
boundary conditions. However, if the fluxes near the source are mutually
BPS with the source, then there is no solution.

To strengthen our results we have considered the possibility that
the sources are not fully localised as a delta-function but instead
have some remaining profile away from the pole on which the source
is placed. Our results show that the solution is uniquely the fully
smeared solution, which is not dependent on any assumptions of
boundary conditions. The fact that there exists no source profile
that is even a bit lumped around the poles makes it unlikely that 
solutions in the fully localised case exist.

This is in strong contrast with the BPS solutions given by the GKP
backgrounds and their T-duals, since there one can find solutions to
the equations of motion for an arbitrary source profile. The only fields
that depend on the specific form of the source are the warp factor
and the $F_5$-flux.

\section{Discussion}\label{Discussion}

One can regard the existence of flux vacua as a consequence of a
balancing of various forces coming from localised sources,
fluxes and curvature. When flux vacua are established using smeared
sources, one has to make sure that the balancing of the forces does not rely
on the sources being smeared out. If this is not satisfied, the solutions could not be trusted,
since the balancing of the forces would be lost for the localised sources, and
the solutions become non-static. Especially for non-BPS
situations one might expect problems since one generically combines
ingredients that are mutually non-BPS and therefore exert non-zero
forces on each other. It is conceivable that smearing sources can undo
such forces and give rise to fake solutions, which exist only due to
the smearing. A simple example with mutually
non-BPS ingredients is an anti-$D3$-brane in an ISD background, or
its T-dual version with an anti-$D6$-brane, considered in this paper in
detail.

The localised version of such a smeared non-BPS solution would lead to brane-like sources surrounded by flux that,
on the average, has the wrong sign to be mutually BPS with the source at a global level.
 However, if the flux changes its BPS-type near the branes so as to be mutually BPS with them in their vicinity,
the balance of forces could be restored at least in the region near the branes. The question then is whether such a
gradient in the flux type can be sustained and produce a static global solution.

We have investigated this using truly delta-like brane sources and
found global obstructions for the existence of a solution, as long as
arbitrarily close to the source the fluxes are mutually BPS with the sources.
We furthermore investigated the existence of solutions with regularised brane
sources and showed that the fully smeared solutions are unique,
thereby confirming the conjecture made in \cite{Blaback:2010sj}. The
reason we investigated regularised sources as well is that they
should tell us about what happens in the localised limit. Intuitively this can be
understood by considering a truly localised brane whose profile is
then a tiny bit smeared. We find in our setup that the entire background flux gets drawn into
the brane region, which can be viewed as the physical reason for the complete breakdown of the solution upon localisation of the sources.

Our work is similar in spirit to the investigations of the backreaction
of anti-$D3$-branes in throat geometries (see e.g.
\cite{DeWolfe:2008zy, McGuirk:2009xx, Bena:2009xk, Bena:2010gs,
Baumann:2010sx, Dymarsky:2011pm, Bena:2011hz } and references therein) but
is different in the sense that we do not break supersymmetry perturbatively
by just a few anti-branes. Instead we have as many SUSY-breaking
branes as there is background flux. However, as we show in the
appendix we can also find smeared solutions with a tunable amount of
SUSY-breaking branes, such that we can be arbitrarily close to the
BPS no-scale Minkowski solutions of \cite{Giddings:2001yu,Blaback:2010sj}.
We study the localisation of these solutions in \cite{toappear}, where we
also discuss the boundary conditions at the sources in more detail.

Our results have particular relevance for attempts to find
well-controlled de Sitter vacua at leading order in the $g_s$ and
$\alpha^{\prime}$ expansion \cite{Hertzberg:2007wc}. Such classical
de Sitter compactifications are surprisingly difficult to engineer
due to a number of simple no-go theorems. The simplest way around
these no-go theorems seems to require the inclusion of orientifold
planes in combination with compact spaces of negative integrated
scalar curvature \cite{Silverstein:2007ac, Haque:2008jz,Wrase:2010ew}. In the
limit of smeared $O$-planes on a few negatively curved manifolds,
classical de Sitter solutions were indeed identified in
\cite{Flauger:2008ad,Caviezel:2008tf,Danielsson:2009ff,Caviezel:2009tu,
Danielsson:2010bc, Danielsson:2011au}\footnote{These solutions
correspond to critical points of lower-dimensional supergravity
theories. But there also exist examples which breaks supersymmetry
at the level of the compactification
\cite{Dong:2010pm, Andriot:2010ju}.}. The known de Sitter models of
this type are still not satisfactory, as they all have a tachyonic
instability (i.e. they correspond to saddle points rather than local
minima in the effective potential), and they suffer from reliability
issues when flux quantization is taken into account
\cite{Danielsson:2011au}. However, a more basic issue with
compactifications on negatively curved spaces was raised in
\cite{Douglas:2010rt}. There it was pointed out that in absence of
warping and noticeable quantum corrections, negative curvature
spaces require the presence of a continuous distribution of negative
energy density as given by a smeared $O$-plane. Upon localisation of
the $O$-planes, however, an everywhere negatively curved compact space
can only be maintained when there are strong warping effects
everywhere or the classical approximation breaks down. As the
localisation of a smeared source does induce a nontrivial warp
factor, one can potentially maintain negative
internal curvature, and indeed in a BPS-like Minkowski
configuration, one can explicitly show that this is possible
\cite{Blaback:2010sj}. The importance of our result lies in the
observation that in non-BPS cases, such as de Sitter solutions, a
localised solution may simply not exist after all. Note also
that our results do not depend on the size of the compact space, which
clearly shows that localisation effects in flux compactifications
are important for arbitrarily large internal spaces.

It should be emphasised that our results not only cast doubts on the
validity of classical de Sitter vacua that have been constructed in
the limit of smeared sources, but that, more generally, non-BPS
configurations supported by smeared $D$-branes or $O$-planes may not
have reliable localized counterparts, even when they are
perturbatively stable in the usual gauged supergravity truncation,
involve $AdS$-spaces or compact spaces of positive curvature.

\section*{Acknowledgements}
We like to thank Iosif Bena, I\~naki Garc\'ia Etxebarria, Ben
Heidenreich, Fernando Marchesano, Liam McAllister, Paul McGuirck,
Diederik Roest, Gary Shiu, Yoske Sumitomo and Bret Underwood for
useful discussions. U.D. is supported by the Swedish Research
Council (VR) and the G\"oran Gustafsson Foundation. D.J. and M.Z.
are supported by the German Research Foundation (DFG) within the
Emmy Noether Program (Grant number ZA 279/1-2) and the Cluster of
Excellence ``QUEST''. T.V.R. is supported by the G\"oran Gustafsson
Foundation. T.W. is supported by the Alfred P. Sloan Foundation and
by the NSF under grant PHY-0757868. This research was supported in
part by the National Science Foundation under Grant No. NSF
PHY05-51164. T.W. thanks the Hong Kong Institute for Advanced Study
for hospitality.

\appendix

\section{Smeared non-BPS solutions for general $d$ and $\eta$}
\label{app_solutions}

Let us consider a combination of smeared $p$-brane sources that can
be a general mixture of $Dp$-, anti-$Dp$-, $Op$- and anti-$Op$-planes.
This way we can get arbitrarily close to the BPS point by taking an
arbitrarily small amount of anti-branes compared to the background
$O$-planes. This makes sense in the supergravity picture where we do
not bother about charge quantisation. Since the question we address
is a supergravity question (the existence of localised solutions),
we settle with this imprecise state of affairs. Because this
mixture might be non-BPS we allow a resulting charge $Q$ and tension
$T$ that might have different magnitudes, i.e,
\begin{equation}\label{eta}
T=\eta Q\,,
\end{equation}
where $\eta$ measures how far we are from a BPS situation. For
example $\eta=1$ with $Q<0$ means a BPS mixture of (anti-)$Op$ and
(anti-)$Dp$ sources such that the orientifold tension dominates. Or
when $\eta=-1$ and $Q < 0$ this corresponds to a net anti-$Dp$
charge and tension. We furthermore point out that there also exist
tachyonic de Sitter solutions with $\eta<-1$ when we allow $Op$- and
anti-$Op$-planes at the same time. It is not immediately clear to us
whether this can be done consistently in the context of flux
compactifications, but examples with $Op$- and anti-$Op$-planes
appeared in \cite{Kachru:1999ed}.

\subsection{The ansatz}
We look for solutions with constant dilaton, $\phi_0$, and an
internal space that is a direct product of two spaces
\begin{align}
& \d s_{10}^2 =\d s_{p+1}^2 + \d\Sigma_3^2 +\d\Sigma_{6-p}^2\,,\\
& H = h\star_3 1\,,\\
& F_{6-p}=(-1)^p\e^{-\tfrac{p+1}{4}\phi_0}\kappa \star_{9-p} H\,,
\end{align}
where $\d s^2_{p+1}$ is the metric of the external space, which we assume
to be maximally symmetric (AdS, Mink, or dS), and $\d\Sigma_3^2$ and $\d\Sigma_{6-p}^2$
are metrics on Euclidean $3$-, resp. $(6-p)$-dimensional Einstein
spaces, sofar left unspecified. The parameter $\kappa$ is assumed
constant. E.g., for $p=3$ we have that $\kappa=+1$ corresponds to ISD
fluxes and $\kappa=-1$ to IASD fluxes. Since $H$ fills the
$\Sigma_3$ space, the duality relation between $F_{6-p}$ and $H$
implies that $F_{6-p}$ fills the $\Sigma_{6-p}$ space, so these are
truly simple ans\"atze.

Let us now go through all equations of motion. The $H$ and $F_{6-p}$
Bianchi identities are trivially satisfied whereas the $F_{8-p}$
Bianchi identity implies the tadpole condition\footnote{In this appendix we set the volume of the internal space $V=1$ or equivalently take $Q$ and $T$ to denote the charge/tension densities.}
\begin{align}\label{tadpole}
& Q = -\e^{-\tfrac{p+1}{4}\phi_0}\kappa h^2\,.
\end{align}
Furthermore, the $H$, $F_{6-p}$ and $F_{8-p}$ equations of motion
are also satisfied. The dilaton equation, on the other hand implies
a second-order polynomial for $\kappa$, which allows two solutions
\begin{equation}\label{kappa}
\kappa_{\pm}=\frac{(p-3)\eta\pm\sqrt{(p-3)^2\eta^2 +
8(p-1)}}{2(p-1)}.
\end{equation}
This implies that we always have two branches of solutions
corresponding to $\kappa_+$ and $\kappa_-$.

Let us analyse the internal Einstein equation along the first space
$\Sigma_3$. We find
\begin{equation}
(R_3)_{ab}=\frac{\e^{-\phi_0}}{16}h^2\Bigl( 6 - \kappa \eta(p+1) -
(5-p)\kappa^2\Bigr)g^{(3)}_{ab},
\end{equation}
which fixes the curvature of the Einstein space $\Sigma_3$.
Similarly we find from the remaining directions in the internal
Einstein equation
\begin{equation}
(R_{6-p})_{ab}=-\frac{\e^{-\phi_0}}{16}h^2\Bigl(2 +
(p+1)\eta\kappa -(p+3)\kappa^2\Bigr)g^{(6-p)}_{ab}\,.
\end{equation}
The external Einstein equation sets the value of the cosmological
constant in the external dimensions
\begin{equation}
R_{p+1} = -\tfrac{1}{8}(p+1)\e^{-\phi_0}h^2\Bigl(1 +
\tfrac{5-p}{2}\kappa^2 - \tfrac{(7-p)}{2}\eta\kappa \Bigr)\,.
\end{equation}

Let us now analyse all the possible solutions, starting with
recovering the BPS and non-BPS solutions of \cite{Blaback:2010sj},
where $\eta=\pm 1$.

\subsection{The solutions with $\eta=\pm1$}
The solutions for $\kappa$ are  (\ref{kappa})
\begin{equation}
\eta=\pm 1\qquad \rightarrow \qquad \kappa = \eta \,\,\,,\qquad
\kappa=-\eta\frac{2}{p-1}\,.
\end{equation}
From equation (\ref{tadpole}) this determines the sign of the charge
and from that we find the sign of the tension using (\ref{eta}). For
the "BPS" values, $\kappa=\eta=\pm 1$, we find that the tension is
necessarily negative
\begin{equation}
T(\kappa=\eta) = -h^2\e^{-\tfrac{p+1}{4}\phi_0},
\end{equation}
 and that the solution is Minkowski and the
internal spaces $\Sigma_3, \Sigma_{6-p}$ are flat. For $p=3$ this is
the smeared GKP solution, built from $O3$-planes and ISD flux when
$\eta=1$ and anti-$O3$-planes and IASD flux when $\eta=-1$.

For the non-BPS values $\kappa= -\tfrac{2}{p-1}\eta$ we find
\begin{align}
 R_{p+1}        &= -\tfrac{(p+1)^2}{2(p-1)^2}\e^{-\phi_0}h^2 \,,\\
 (R_{3})_{ab}   &= \tfrac{(p+1)(p-2)}{2(p-1)^2}\e^{-\phi_0}h^2\, g^{(3)}_{ab}\,,\\
 (R_{6-p})_{ab} &=
\tfrac{(p+1)}{2(p-1)^2}\e^{-\phi_0}\,h^2\,\,g^{(6-p)}_{ab} \,.
\end{align}
Hence these are AdS solutions with positively curved internal
Einstein spaces, which we can for instance take to be
spheres\footnote{The case $p=2$ is exceptional because the 3-space
$\Sigma_3$ filled with $H$-flux is flat.}. Combining equations
(\ref{eta}, \ref{tadpole}) we find the total tension to be positive
\begin{equation}
T(\kappa=\tfrac{-2}{p-1}\eta)=\tfrac{2}{p-1}h^2\e^{-\tfrac{p+1}{4}\phi_0}\,,
\end{equation}
corresponding to net $D$-brane tension.

The solutions with $\eta=\pm 1$ are special in the sense that
supersymmetry is not broken by the sources, but by choosing fluxes
that are not mutually BPS with the branes. This implies that, under
certain conditions, these AdS solutions are expected to be critical
points of half-maximal gauged supergravities. This seems the case
when we take as internal Einstein spaces the $S^3\times S^{6-p}$ (or
$T^3\times S^4$ when p=2). This is because
sphere-reductions, when consistent, are not expected to break the
supersymmetry of the ten-dimensional action, only the sources break
half of the supersymmetries. Especially for $p=3, 5$ and $6$ we are
bound to have a description in terms of half maximal gauged
supergravity since $S^3\times S^{6-p}$ is then a group manifold and
dimensionally reducing on a group manifold is consistent
\cite{Scherk:1979zr}.

\subsection{The solutions with general $\eta$}
Let us now consider general $\eta$. We find
\begin{align}
R_{p+1}        &= -\tfrac{(p+1)}{2(p-1)}\e^{-\phi_0}\,h^2\,\bigl(1-\eta\kappa\bigr)\,,\\
(R_{3})_{ab}   &= \tfrac{(p-2)}{2(p-1)}\e^{-\phi_0}\,h^2\,\Bigl(1-\eta\kappa\Bigr)g^{(3)}_{ab}\,,\\
(R_{6-p})_{ab} &=
\tfrac{1}{2(p-1)}\e^{-\phi_0}\,h^2\,\Bigl(1-\eta\kappa
\Bigr)g^{(6-p)}_{ab}\,,
\end{align}
with the total tension of the sources given by
\begin{equation}\label{tension}
T = -\e^{-\tfrac{p+1}{4}\phi_0}\eta\kappa h^2\,.
\end{equation}
From these expressions we notice the following structure
\begin{itemize}
\item For the range $\kappa\eta<1$ we have non-BPS AdS vacua with positively
curved internal Einstein spaces.
\item For the range $\kappa\eta=1$ we have the BPS Minkowski solutions with
flat internal spaces and net orientifold sources.
\item For the range $\kappa\eta>1$ we have de Sitter solutions with
negatively curved internal Einstein spaces and net orientifold
sources.
\end{itemize}
By using the dependence of $\kappa$ on $\eta$ we can verify that
$\kappa\eta>1$ necessarily implies $\eta^2>1$, which implies that
the de Sitter solutions are only possible for combinations of $Op$-
and anti-$Op$-planes since such combinations have more negative
tension than charge. The Minkowski turning point $\eta\kappa=1$
implies $\eta^2=1$ and is thus only possible for BPS sources.

In order to understand the perturbative stability of the solution we
also computed the masses of 3 universal scalar fields: the dilaton
$\phi$, the volume modulus $\varphi_1$, and the modulus $\varphi_2$
that measures the relative sizes of the two internal Einstein spaces
(and does not exist for the $p=6$ solution). We found that the AdS
solutions are stable within this subspace of degrees of freedom and
the dS solutions are not.

\bibliography{groups}

\bibliographystyle{utphysmodb}

\end{document}